\documentclass[conference]{IEEEtran}
\IEEEoverridecommandlockouts
\usepackage{cite}
\usepackage{amsmath,amsthm,amsfonts,amssymb}
\usepackage{graphicx}
\usepackage{textcomp}
\usepackage{subfig}
\usepackage{bm}
\usepackage{bbm}
\usepackage{url}
\usepackage{array}
\usepackage{color}
\usepackage{multirow}
\usepackage{booktabs}
\usepackage[table,xcdraw]{xcolor}
\usepackage{enumerate}
\usepackage{epstopdf}
\usepackage{tabularx}
\usepackage{algorithm}
\usepackage[english]{babel}
\usepackage{algpseudocode}
\usepackage[misc]{ifsym}

\begin{document}
\title{}
\title{\makebox[\linewidth]{\parbox{\dimexpr\textwidth+0cm\relax}{\centering LLM-Assisted Multi-Teacher Continual Learning for Visual Question Answering in Robotic Surgery}}}
\author{
    \IEEEauthorblockN{Yuyang Du{\small$^\dagger$}, Kexin Chen{\small$^\dagger$}, Yue Zhan{\small$^\dagger$},  Chang Han Low, Tao You, Mobarakol Islam, Ziyu Guo,\\ Yueming Jin$^*$,
    Guangyong Chen, Pheng-Ann Heng}  \vspace{-6.5em}
    \IEEEauthorblockA{\thanks{Y. Du is with the Department of Information Engineering, the Chinese University of Hong Kong (CUHK), HKSAR, China. K. Chen, Z. Guo, and P. A. Heng are with the Department of Computer Science Engineering, CUHK, HKSAR, China. P. A. Heng is also with Institute of Medical Intelligence and XR, CUHK, HKSAR, China. Y. Zhan is with the Department of Electrical and Electronic Engineering, the University of Hong Kong (HKU), Hong Kong SAR, China. C. H. Low, T. You and Y. Jin are with Jin Lab, National University of Singapore (NUS), Singapore. M. Islam is with the Department of Medical Physics and Biomedical Engineering, University College London (UCL), London, UK. G. Chen is with Zhejiang Lab, Zhejiang, China.}} 
    \IEEEauthorblockA{\thanks{The work was partially supported by the Research Grants Council of the Hong Kong Special Administrative Region, China (Project Number: T45-401/22-N), by the Hong Kong Innovation and Technology Fund (Project Number: GHP/080/20SZ), by the Regional Joint Fund of Guangdong (Guangdong-Hong Kong-Macao Research Team Project) under Grant 2021B1515130003, by the Guangdong-Hong Kong-Macao Joint Laboratory of Human-Machine Intelligence-Synergy Systems, and by the Ministry of Education Tier 1 Start-up grant, NUS, Singapore (WBS: A-8001267-01-00).}} 
    \IEEEauthorblockA{\thanks{Code and data availability: https://github.com/yuyangdu01/LLM-CL-VQA.}}
    \IEEEauthorblockA{\thanks{$^*${Yueming Jin (ymjin@nus.edu.sg) is the corresponding author}.}}
    \IEEEauthorblockA{\thanks{$^\dagger$Y. Du, K. Chen, and Y. Zhan are equal contributors.}} 
}
\maketitle
\begin{abstract}
Visual question answering (VQA) can be fundamentally crucial for promoting robotic-assisted surgical education. In practice, the needs of trainees are constantly evolving, such as learning more surgical types and adapting to new surgical instruments/techniques. Therefore, continually updating the VQA system by a sequential data stream from multiple resources is demanded in robotic surgery to address new tasks. In surgical scenarios, the privacy issue of patient data often restricts the availability of old data when updating the model, necessitating an exemplar-free continual learning (CL) setup. However, prior studies overlooked two vital problems of the surgical domain: i) large domain shifts from diverse surgical operations collected from multiple departments or clinical centers, and ii) severe data imbalance arising from the uneven presence of surgical instruments or activities during surgical procedures. This paper proposes to address these two problems with a multimodal large language model (LLM) and an adaptive weight assignment methodology. We first develop a new multi-teacher CL framework that leverages a multimodal LLM as the additional teacher. The strong generalization ability of the LLM can bridge the knowledge gap when domain shifts and data imbalances occur. We then put forth a novel data processing method that transforms complex LLM embeddings into logits compatible with our CL framework. We also design an adaptive weight assignment approach that balances the generalization ability of the LLM and the domain expertise of the old CL model. Finally, we construct a new dataset for surgical VQA tasks. Extensive experimental results demonstrate the superiority of our method to other advanced CL models.
\end{abstract}


\section{Introduction}\label{sec-I}
Robot-assisted surgery (RAS) has garnered growing importance in recent years due to its distinct advantages such as stability, precision, and ultra efficiency with minimal human effort required \cite{Ref003, Ref005, Ref045}. However, teaching junior medical students operations related to RAS is time-consuming, as they may have many questions emerging during the training process. Given the critical nature of surgical operations, these questions should be clearly addressed. Yet, expert surgeons may not always provide feedback promptly, as they are often overloaded with heavy clinical and academic workloads \cite{Ref006, Ref007, Ref008}. Therefore, developing visual question answering (VQA) models from expert demonstration videos attracts a surge of research interest \cite{Ref035, Ref036, Ref037}, as a VQA model not only liberates expert surgeons from repeating teaching procedures but also facilitates personalized learning experience of trainees. Moreover, VQA can be incorporated into embodiment to endow the system with advanced cognitive ability to comprehend and interpret surgical scenes, which lays the groundwork for the next generation of intelligent and autonomous surgical robots.

In surgical VQA, the needs of trainees are constantly evolving, such as learning more surgical types and adapting to different robotic systems. Meanwhile, new techniques and instruments are consistently introduced to enhance patient care. These introduce new environments (new surgical scenes) and new question-and-answer sets, leading to various new VQA tasks. Continual Learning (CL) ability on new tasks is highly crucial in developing advanced VQA systems. Conventional machine learning methods are likely to suffer from significant performance degradation on prior tasks when acquiring new knowledge, also known as catastrophic forgetting \cite{Ref009}. Pioneering works in the medical domain addressed this problem by adapting CL algorithms of general domain \cite{RefA2, RefA1}. For example, \cite{RefA2} developed a replay-oriented CL algorithm for medical image analysis. These studies are exemplar-based, where old patient data is accessible when updating the model. However, in realistic surgical VQA, it is preferable to incorporate new knowledge from multiple sources with more data, such as multiple clinical centers; due to high storage costs, data privacy restrictions, and licensing issues across different centers, exemplar-based methods are not practical.

Some recent works also focus on CL studies under the data privacy restriction. \cite{RefA4} adapted non-exemplar-based CL algorithms like Learning without Forgetting (LwF) \cite{Ref001} and Elastic Weight Consolidation (EWC) \cite{Ref002} in medical tasks. \cite{RefA3} introduced CL algorithms in instrument and tissue localization tasks alongside VQA, with a special focus on overlapping classes that appear in both the new dataset and the old one. However, we identify two main properties in the robotic surgery domain that have been largely overlooked in the existing literature, leading to their unreliable performance when tackling complex surgical VQA tasks. The two properties include \textit{large domain shift} and \textit{severe data imbalance}. 

\textbf{Domain Shift}: When updating a CL model under the classic teacher-student framework, the model trained on old data (teacher) serves as an important reference for the new model. The new model (student) utilizes the logits of the old model during training, allowing the retention of the former's knowledge without revisiting any old data \cite{Ref027}. If a teacher encounters an entirely unfamiliar task in the new training data, it just provides random guesses for the student's reference \cite{Ref028, Ref029}, referred to as the domain shift. In robotic surgical applications, such domain shift is notably prevalent: surgical scenes from different types of surgeries show a large variety of appearances, even within specific categories like abdominal surgery, due to differences in instruments and surgical actions. For example, data for the surgery performed on a liver largely diverges from that on a kidney. The data collected from multiple sources may further increase this problem, given variations in operation protocols and robotic systems. This severe domain shift commonly exists in surgical datasets and significantly reduces the performance of the CL model, as the student learns nothing but the teacher's random guesses.

\textbf{Data Imbalance}: In practical surgeries, some actions or instruments are less frequently presented. For example, the action of tissue manipulation frequently appears in many surgical datasets, given that most operations involve tissue interaction. Cutting action is also widely mentioned, especially in the dataset focusing on nephrectomy, while stapling is rarely mentioned in the same dataset because stapling is generally performed only after vessel cutting, a niche operation inside the whole nephrectomy. In training data, if some classes have significantly fewer examples than others, we identify this as data imbalance \cite{Ref030, Ref031}. Prior CL algorithms \cite{Ref001, Ref002, RefA1, RefA2, RefA3, RefA4} did not particularly address imbalanced data, making the model inadequately trained for minor classes.

This paper leverages the strong generalization ability of multimodal large language models (LLMs) to overcome the knowledge limitations imposed by the two problems. Multimodal LLMs are generative AI models trained on vast amounts of image and text data collected across various domains \cite{Ref010, Ref011, Ref012, Ref043}. Recently, they have received increasing interest in various research fields due to their ability to answer questions across diverse domains. Importantly, state-of-the-art LLMs have demonstrated promising competency in answering questions for the medical domain \cite{Ref013, Ref014}. This motivates us to utilize LLM-generated answers to fill the knowledge gap when training data exhibits new knowledge unfamiliar to the teacher model or is extremely imbalanced. An intuitive explanation is that, when faced with knowledge outside the teacher's expertise, leveraging the robust medical understanding of an LLM is far superior to relying on the teacher model's poorly trained knowledge (for imbalanced data) or even random guesses (for domain shifts).

Apart from the LLM-aided multi-teacher CL framework, this paper also devises an adaptive weight assignment approach to balance insights from the LLM's general knowledge and the conventional teacher's domain-specific expertise. With adaptive weights, the student training process reaches an ideal state: it takes more insights from the conventional teacher when the knowledge belongs to a familiar domain and is well-trained with rich data; otherwise, it relies more on the LLM. 

Furthermore, we construct a new surgical VQA dataset to validate our innovations in real-world surgical settings. We developed a novel GPT-based QA pair generation method when building the new VQA dataset. We apply in-context learning (ICL) \cite{Ref015, Ref032, Ref044a, Ref044b, Ref033}, an advanced few-shot learning method, for better analysis of text descriptions for a clinical image.

\begin{figure*}[htbp]
  \centering
  \includegraphics[width=0.775\textwidth]{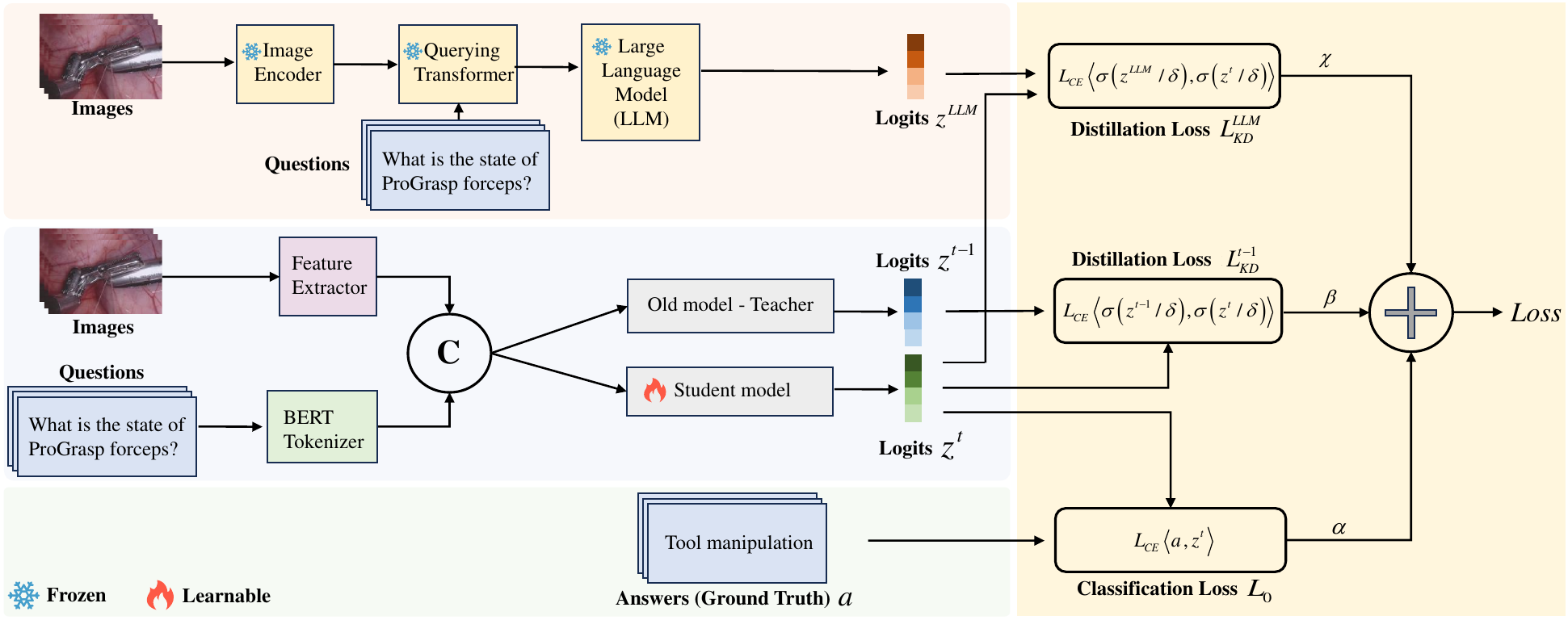}\\
  \captionsetup{font={small}}
  \caption{The proposed LLM-assisted multi-teacher CL framework. The model is used to process bimodal input (text and image) and provide predictions for the VQA task. The proposed weight adaption scheme (highlighted in the light orange zone) is designed to trade off the general knowledge of the LLM and the medical expertise of the previous CL model. The frozen LLM is highlighted in light red, while the conventional teacher and student models are highlighted in light blue. Ground truth lies in the light green zone.}\label{plot3}
  \vspace{-1.5mm}
\end{figure*}
\section{Proposed Method}\label{sec-II}
\subsection{Preliminaries}
\textbf{Problem Formulation and Notations:} 
Let us assume a CL process with $\tau$ time period, in which $t\in \left\{ 1,...,\tau  \right\}$ denotes a time period of the process. We denote the training dataset in time $t$ by ${{D}_{t}}$, wherein element ${{d}_{t,i}}\in {{D}_{t}}$ represents the ${{i}^{th}}$ training sample for time $t$. Each training sample ${{d}_{t,i}}$ contains a surgical frame and several closely associated clinical questions. We denote the classes appearing in ${{D}_{t}}$ by ${{C}_{t}}$, wherein element ${{c}_{t,j}}$ represent the ${{j}^{th}}$ class appearing in ${{D}_{t}}$. If class ${{c}_{t,j}}$ frequently appears in ${{D}_{t}}$ but never appears in previous training datasets (i.e., ${{D}_{t-1}}$, ${{D}_{t-2}}$,…, ${{D}_{1}}$), we say domain shift happens at class ${{c}_{t,j}}$. Further, if the appearing frequency of class ${{c}_{t,j}}$ is significantly smaller than other classes in ${{D}_{t}}$, then data imbalance happens at class ${{c}_{t,j}}$.

\textbf{Distillation Loss in Continual Learning:} In CL, knowledge is obtained from an infinite stream of data, with the goal of gradually updating the model with new data and avoiding forgetting the knowledge learned in old data. Knowledge distillation (KD) is an efficient approach for retaining the knowledge of the old model and addressing the catastrophic forgetting problem without revisiting the previous training data \cite{Ref026}. There are three categories of KD: response-based KD, feature-based KD, and relation-based KD. In this paper, we focus on response-based KD because it is flexible and allows student and teacher models of different network types \cite{Ref034}.

The distillation loss of the KD process is formulated as
\begin{equation}\label{eq001}
\small
\begin{aligned}
{{L}_{KD}}={{L}_{CE}}\left\langle \sigma \left( {{{z}^{T}}}/{\delta }\; \right),\sigma \left( {{{z}^{S}}}/{\delta }\; \right) \right\rangle
\end{aligned}
\normalsize
\end{equation}
where ${{L}_{CE}}\left\langle \cdot ,\cdot  \right\rangle$ denotes the cross-entropy loss; $\sigma \left( \cdot  \right)$ denotes the softmax function; ${{z}^{T}}$ and ${{z}^{S}}$ denote the output logits of the teacher and the student model, respectively; $\delta$ is a temperature hyperparameter that controls the softness of the probability distributions. With $\delta =1$, we get the standard softmax function, and as $\delta$ increases, the produced probability distribution becomes softer, providing more information such as which class is more similar to the predicted class.

\subsection{Multi-teacher CL Framework with LLM} 
To overcome domain shifts and data imbalances, we introduce an additional multimodal LLM teacher with a strong generalization ability for better knowledge transfer. When the model in time $t$ is unfamiliar with the knowledge in time $t+1$, the LLM teacher will guide the student to learn from a more reasonable knowledge source.

Our multi-teacher CL framework is presented in Fig. \ref{plot3}. As the figure shows, the general loss function $L$ is written as
\begin{equation}\label{eq002}
\small
\begin{aligned}
L=\alpha {{L}_{0}}+\beta L_{KD}^{t-1}+\chi L_{KD}^{LLM}
\end{aligned}
\normalsize
\end{equation}
where $\alpha$, $\beta$, and $\chi$ are normalized adaptive weights with a sum of one (say Section \ref{sec-II}-D for the weight assignment scheme); ${{L}_{0}}$ is the cross-entropy loss supervised by the hard labels; $L_{KD}^{t-1}$ is the KD loss between the new CL model trained in time $t$ (i.e., the student model) and the old CL model trained in time $t-1$ (i.e., the conventional teacher model); $L_{KD}^{LLM}$ is the KD loss between the student model and the LLM teacher. In this paper, we denote the logits of the student model, the old teacher model, and the LLM teacher mode by ${{z}^{t}}$, ${{z}^{t-1}}$, and ${{z}^{LLM}}$, respectively. And we write $L_{KD}^{t-1}$ and $L_{KD}^{LLM}$ as
\begin{equation}\label{eq003}
\small
\begin{aligned}
L_{KD}^{t-1}={{L}_{CE}}\left\langle \sigma \left( {{{z}^{t-1}}}/{\delta }\; \right),\sigma \left( {{{z}^{t}}}/{\delta }\; \right) \right\rangle
\end{aligned}
\normalsize
\end{equation}
\begin{equation} \label{eq004}
\small
\begin{aligned}
L_{KD}^{LLM}={{L}_{CE}}\left\langle \sigma \left( {{{z}^{LLM}}}/{\delta }\; \right),\sigma \left( {{{z}^{t}}}/{\delta }\; \right) \right\rangle 
\end{aligned}
\normalsize
\end{equation}

Since the LLM teacher is implemented in a complex transformer network that is significantly different from the conventional teacher model and the student model, some transformation works are needed to obtain the logits from the embeddings. We next give details about the selected LLM and the embedding-logit transformation as follows.

We select InstructBLIP \cite{Ref016}, an open-source multimodal LLM with vision and language ability, as the LLM teacher. There are three components in InstructBLIP: one image encoder that deals with the image input, one text-in-text-out LLM to manage the output, and an image-text transformer to bridge the two modules. Thanks to the modular architectural design, InstructBLIP is highly flexible, and we can quickly adapt a wide range of text-to-text LLMs for implementation. Without loss of generality, we utilize FlanT5 \cite{Ref017}, an instruction-tuned model based on Transformer T5, as the text-in-text-out LLM.


The embeddings we obtained from the last fully connected layer of FlanT5 is a self-attention matrix demoted by ${{e}_{LLM}}$. The size of ${{e}_{LLM}}$ is $N\times \left( M+1 \right)\times P$, where $N$ is the number of classes in time $t$, i.e., the cardinality of ${{C}_{t}}$; $P$ is the LLM's vocabulary; and $M$ is the number of tokens we used to represent each class label.\footnote{The second dimension of the matrix is $M+1$ rather than $M$ because we add one additional pause token for each word to indicate the end of a word.} As presented in Eq. (\ref{eq004}), the desired logits ${{z}^{LLM}}$ should have a size of $N\times 1$. We now illustrate how we transform the $N\times \left( M+1 \right)\times P$ embeddings into the $N\times 1$ logits.
\begin{figure}[htbp]
  \centering
  \includegraphics[width=0.35\textwidth]{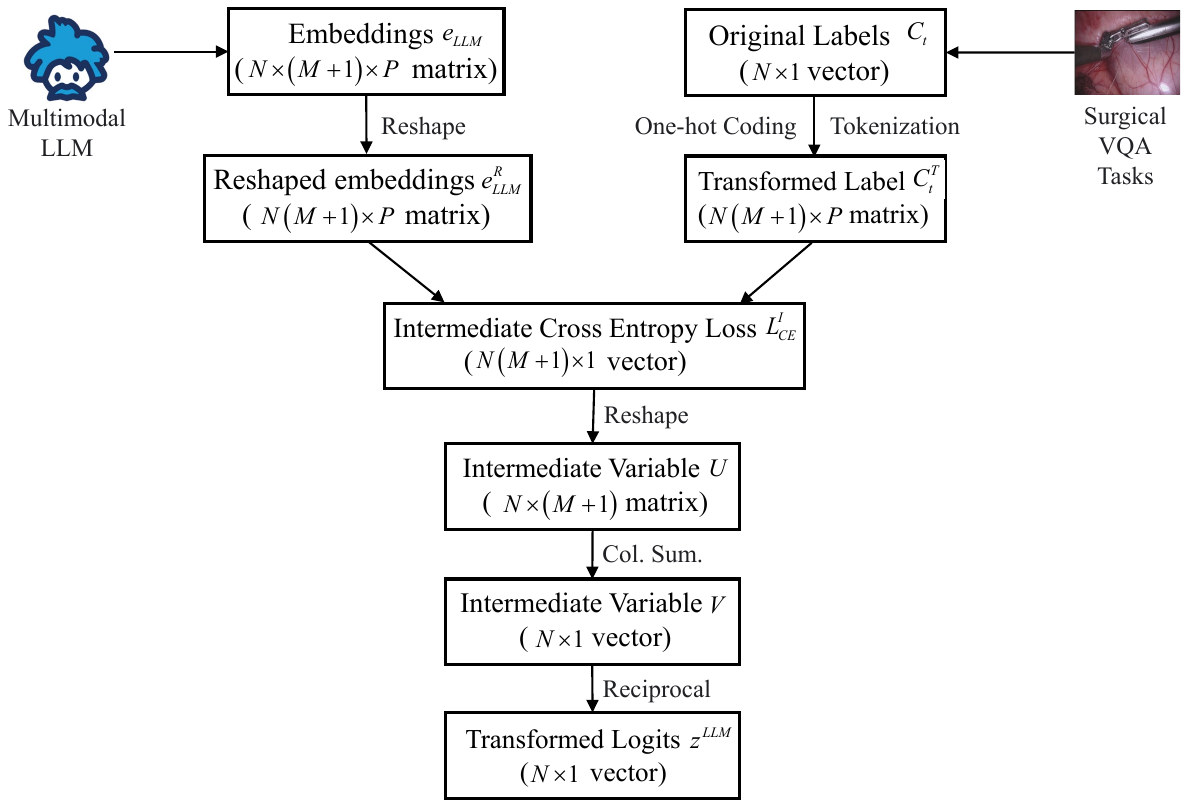}\\
  \captionsetup{font={small}}
  \caption{The workflow of logits transformation after embeddings.}
  \vspace{-1em}
  \label{plot4}
\end{figure}

As in Fig. \ref{plot4}, we first reshape the extracted embeddings and obtain a new matrix $e_{LLM}^{R}$. We conduct one-hot encoding and tokenization on the classification label set ${{C}^{t}}$. Note that $e_{LLM}^{R}$ and ${{C}^{T}}$ have the size of $N\left( M+1 \right)\times P$. We then calculate the cross-entropy loss between $e_{LLM}^{R}$ and ${{C}^{T}}$ by
\begin{equation}\label{eq005}
\small
\begin{aligned}
L_{CE}^{I}\left( i \right)={{L}_{CE}}\left\langle e_{LLM}^{R}(i),{{C}^{T}}(i) \right\rangle 
\end{aligned}
\normalsize
\end{equation}

The resulting vector $L_{CE}^{I}$ has a size of $N\left( M+1 \right)\times 1$. We next reshape the vector to a $N\times \left( M+1 \right)$ matrix and conduct column summation to obtain the cross entropy corresponding to each label. In the resulting $N\times 1$ vector $V$, element $V(j)$ represents the expected loss of label $c_{t,j}$, which is negatively related to the probability of label $c_{t,j}$ being selected as the final classification result. While in the desired logits vector ${{z}^{LLM}}$, element ${{z}^{LLM}}\left(j\right)$ should be positively related to the classification probability of $c_{t,j}$ so that $\sigma \left( {{{z}^{LLM}}}/{\delta }\; \right)$, the output of the softmax layer, can be a probability vector. Therefore, we need to further transform vector $V$ to make its elements positively related to the classification probability. One possible way is to inverse the elements in $V$, i.e., ${{z}^{LLM}}\left( j \right)={1}/{V\left(j\right)}$.

\subsection{Adaptive Weight Assignment}
In Eq. (\ref{eq002}), $\alpha$, $\beta$, and $\chi$ denote the weights of ${{L}_{0}}$, $L_{KD}^{t-1}$, and $L_{KD}^{LLM}$, respectively. This paper adaptively adjusts $\beta$ and $\chi$ during the training process with the weight assignment approach described below and sets $\alpha$ as a hyperparameter. Together the three items will jointly control the model optimization.

During the model training at time $t$, the severity of the domain shift of a medical dataset can be assessed by the old model's classification accuracy on the training dataset $D_t$. If the old CL model achieves high accuracy on $D_t$, we infer it possesses familiarity with the knowledge in $D_t$. In contrast, if it has poor accuracy on $D_t$, say worse than the LLM teacher's accuracy, we conclude the old CL model lacks reliable expertise for this training iteration. Given this rationale, we adjust $\beta$ and $\chi$ based on the relative accuracies of the CL teacher and LLM teacher on $D_t$. A greater accuracy differential indicates a more severe domain shift, so more weight should be assigned to the LLM to leverage its generalization ability and general knowledge in the medical domain.

In addition to mitigating domain shift, we aim to address the data imbalance present in the surgical-related dataset via adaptive weight. We observed a unique imbalance in the label distribution of the tested dataset, where some surgical operations and instruments were queried very frequently, while others were rarely mentioned. For instance, in a surgical dataset pertaining to nephrectomies, the action of cutting is frequently referenced, while mentions of stapling are relatively rare because stapling is a specialized procedure performed only after vessel cutting, a niche operation within the broader context of nephrectomy. Although the old CL model may have some knowledge in these seldomly mentioned domains, its knowledge grasp is far from expert due to the scarcity of training data. In such cases, we give the LLM teacher a higher weight to help the training process use its general medical insights to fill in the knowledge gap.

With the above discussion, we see that the assignments of $\beta$ and $\chi$ are jointly decided by domain shifts and data imbalances. Therefore, we express $\beta$ and $\chi$ as
\begin{equation}\label{eq006a}
\small
\begin{aligned}
\beta ={{\theta }_{DS}}{{\beta }_{DS}}+{{\theta }_{DI}}{{\beta }_{DI}}
\end{aligned}
\normalsize
\text{   and   }
\small
\begin{aligned}
\chi ={{\theta }_{DS}}{{\chi }_{DS}}+{{\theta }_{DI}}{{\chi }_{DI}}
\end{aligned}
\normalsize
\end{equation}
where we assume that hyperparameters ${{\theta }_{DS}}$ and ${{\theta }_{DI}}$ satisfy ${{\theta }_{DS}}+{{\theta }_{DI}}=1-\alpha$, reflecting the importance of domain shift (DS) and data imbalance (DI) in our model training, respectively. If we want the model training process focus more on domain shift, we increase ${{\theta }_{DS}}$ and decrease ${{\theta }_{DI}}$; on the other hand, if we want the training process focuses more on data imbalance, we operate conversely. Further, ${{\beta }_{DS}}$ and ${{\chi }_{DS}}$ in Eq. (\ref{eq006a}) denote the weight share of the old CL teacher and the LLM teacher in terms of domain shifts, respectively; ${{\beta }_{DI}}$ and ${{\chi }_{DI}}$ denote the weight share of the two teachers in terms of data imbalance. To have $\beta +\chi =1-\alpha $, we make sure
\begin{equation}\label{eq007a}
\small
\begin{aligned}
{{\beta }_{DS}}+{{\chi }_{DS}}=1
\end{aligned}
\normalsize
\text{   and   }
\small
\begin{aligned}
{{\beta }_{DI}}+{{\chi }_{DI}}=1
\end{aligned}
\normalsize
\end{equation}

We now detail how we assign ${{\beta }_{DS}}$ and ${{\chi }_{DS}}$. We know that ${{\beta }_{DS}}$ and ${{\chi }_{DS}}$ are assigned according to the accuracies of the old CL teacher and the LLM teacher on ${{D}_{t}}$. Given $D_t$, let us denote the classification accuracy of the CL model in time $t-1$ and that of the LLM by $acc_{t-1}$ and $acc_{LLM}$, respectively. To fulfill the constraint in Eq. (\ref{eq007a}), we let ${{\beta }_{DS}}$ and ${{\chi }_{DS}}$ be
\begin{equation}\label{eq008a}
\small
\begin{aligned}
{{\beta }_{DS}}=\frac{ac{{c}_{t-1}}}{ac{{c}_{t-1}}+ac{{c}_{LLM}}}
\end{aligned}
\normalsize
\text{   and   }
\small
\begin{aligned}
{{\chi }_{DS}}=\frac{ac{{c}_{LLM}}}{ac{{c}_{t-1}}+ac{{c}_{LLM}}}
\end{aligned}
\normalsize
\end{equation}
The domain shift problem can be well addressed with the assignment scheme in Eq. (\ref{eq008a}), as the LLM will be given a higher weight when the old CL model is random guessing.

We next introduce how ${{\beta }_{DI}}$ and ${{\chi }_{DI}}$ are decided. In time $t$, we denote the set of previously appeared data by ${{D}_{t,...,1}}$ (i.e., ${{D}_{t,...,1}}={{D}_{t}}\cup {{D}_{t-1}}...\cup {{D}_{1}}$). The ${{k}^{th}}$ class labels in ${{D}_{t,...,1}}$ are denoted by ${{c}_{k}}$, and the appearance time of ${{c}_{k}}$ is denoted by ${{d}_{k}}$. To evaluate the data imbalance in ${{D}_{t,...,1}}$, we introduce the concept of imbalance ratio as in \cite{Ref018}, which is
\begin{equation}\label{eq010}
\small
\begin{aligned}
IR={\max ({{d}_{k}})}/{\min ({{d}_{k}})}
\end{aligned}
\normalsize
\end{equation}
It is important to note that $d_k$ in Eq. (\ref{eq010}) cannot be zero given its definition described above.

Under the constraint in Eq. (\ref{eq007a}), we design ${{\beta }_{DI}}$ and ${{\chi }_{DI}}$ as
\begin{equation} \label{eq011a}
\small
\begin{aligned}
{{\beta }_{DI}}=\frac{1}{1+\log_N IR}
\end{aligned}
\normalsize
\text{   and   }
\small
\begin{aligned}
{{\chi }_{DI}}=\frac{\log_N IR}{1+\log_N IR}
\end{aligned}
\normalsize
\end{equation}
where $N$ is a hyperparameter.

From Eq. (\ref{eq011a}), we can see that when the data is severely imbalanced, we have a large ${{\chi }_{DI}}$ so that the LLM’s general domain knowledge can be well utilized to fill the knowledge gap. The data imbalance problem can be well alleviated. 

Finally, we give the expression of $\beta$ and $\chi$ as follows:
\begin{equation}\label{eq012}
\small
\begin{aligned}
\beta ={{\theta }_{DS}}\frac{ac{{c}_{t-1}}}{ac{{c}_{t-1}}+ac{{c}_{LLM}}}+{{\theta }_{DI}}\frac{1}{1+ \log_N IR}
\end{aligned}
\normalsize
\end{equation}
\begin{equation}\label{eq013}
\small
\begin{aligned}
\chi ={{\theta }_{DS}}\frac{ac{{c}_{LLM}}}{ac{{c}_{t-1}}+ac{{c}_{LLM}}}+{{\theta }_{DI}}\frac{\log_N IR}{1+\log_N IR}
\end{aligned}
\normalsize
\end{equation}

\begin{table*}[htbp]\label{table:1}
\captionsetup{font={small}}
\scriptsize
\caption{Benchmarking experiments (EV17, EV18, and D-V represent EndoVis17, EndoVis18, DAISI-VQA dataset, respectively).}
\vspace{-1em}
\begin{center}
\begin{tabular}
{p{1.75cm}<{\centering}
|p{0.7cm}<{\centering}p{0.7cm}<{\centering}p{0.7cm}<{\centering}
|p{0.7cm}<{\centering}p{0.7cm}<{\centering}p{0.7cm}<{\centering}p{0.7cm}<{\centering}
|p{0.7cm}<{\centering}p{0.7cm}<{\centering}p{0.7cm}<{\centering}
|p{0.7cm}<{\centering}p{0.7cm}<{\centering}p{0.7cm}<{\centering}p{0.7cm}<{\centering}
}
\hline
{\multirow{2}{*}{}} 
&\multicolumn{3}{c|}{Accuracy ($t=1$ to $t=2$)} 
&\multicolumn{4}{c|}{Accuracy ($t=2$ to $t=3$)}
&\multicolumn{3}{c|}{F-Score ($t=1$ to $t=2$)} 
&\multicolumn{4}{c}{F-Score ($t=2$ to $t=3$)}
\\ \cline{2-15}
&EV17 & EV18 & Avg.
&EV17 & EV18 & D-V & Avg.
&EV17 & EV18 & Avg.
&EV17 & EV18 & D-V & Avg.
\\ \hline
FT
& 0.2917 & 0.5905 & \underline{0.4411} & 0.0938 & 0.3286 & 0.7632 & \underline{0.3952} 
& 0.1843 & 0.3806 & \underline{0.2825} & 0.0327 & 0.0982 & 0.8751 & \underline{0.3353} \\  
ER
& 0.5417 & 0.5782 & \underline{0.5599} & 0.5313 & 0.6071 & 0.7544 & \underline{0.6309}
& 0.3344 & 0.3681 & \underline{0.3512} & 0.2784 & 0.3792 & 0.8721 & \underline{0.5099} \\
LwF
& 0.4479 & 0.5309 & \underline{0.4894} & 0.3229 & 0.4124 & 0.6930 & \underline{0.4761}
& 0.3034 & 0.2966 & \underline{0.3000} & 0.1682 & 0.1870 & 0.8419 & \underline{0.3990} \\
Online-EWC 
& 0.4167 & 0.5002 & \underline{0.4584} & 0.0625 & 0.3611 & 0.7368 & \underline{0.3868}
& 0.2276 & 0.2012 & \underline{0.2144} & 0.0362 & 0.1316 & 0.8648 & \underline{0.3442} \\
EWC++ 
& 0.4792 & 0.4680 & \underline{0.4736} & 0.0938 & 0.3734 & 0.7105 & \underline{0.3926}
& 0.2229 & 0.2624 & \underline{0.2427} & 0.0532 & 0.1922 & 0.7293 & \underline{0.3249} \\
\textbf{Our Method} 
& \textbf{0.5104} & \textbf{0.5619} & \underline{\textbf{0.5362}} & \textbf{0.3229} & \textbf{0.4709} & \textbf{0.7368} & \underline{\textbf{0.5102}} 
& \textbf{0.3091} & \textbf{0.3185} & \underline{\textbf{0.3138}} & \textbf{0.2106} & \textbf{0.2556} & \textbf{0.8576} & \underline{\textbf{0.4413}} \\
\hline
\end{tabular}
\end{center}
\normalsize
\vspace{-2mm}
\end{table*}
\begin{table*}[ht]\label{table:2}
\captionsetup{font={small}}
\scriptsize
\caption{Ablation study (EV17, EV18, and D-V represent EndoVis17, EndoVis18, DAISI-VQA dataset, respectively).}
\vspace{-1em}
\begin{center}
\begin{tabular}
{p{1.75cm}<{\centering}
|p{0.7cm}<{\centering}p{0.7cm}<{\centering}p{0.7cm}<{\centering}
|p{0.7cm}<{\centering}p{0.7cm}<{\centering}p{0.7cm}<{\centering}p{0.7cm}<{\centering}
|p{0.7cm}<{\centering}p{0.7cm}<{\centering}p{0.7cm}<{\centering}
|p{0.7cm}<{\centering}p{0.7cm}<{\centering}p{0.7cm}<{\centering}p{0.7cm}<{\centering}
}
\hline
{\multirow{2}{*}{}} 
&\multicolumn{3}{c|}{Accuracy ($t=1$ to $t=2$)} 
&\multicolumn{4}{c|}{Accuracy ($t=2$ to $t=3$)}
&\multicolumn{3}{c|}{F-Score ($t=1$ to $t=2$)} 
&\multicolumn{4}{c}{F-Score ($t=2$ to $t=3$)}
\\ \cline{2-15}
&EV17 & EV18 & Avg.
&EV17 & EV18 & D-V & Avg.
&EV17 & EV18 & Avg.
&EV17 & EV18 & D-V & Avg.
\\ \hline
Scenario 1
& 0.4271 & 0.5677 & \underline{0.4974} & 0.1042 & 0.3337 & 0.7456 & \underline{0.3945}
& 0.2336 & 0.2365 & \underline{0.2351} & 0.0319 & 0.0988 & 0.8673 & \underline{0.3327} \\
Scenario 2 
& 0.4063 & 0.5702 & \underline{0.4882} & 0.0938 & 0.2376 & 0.7632 & \underline{0.3648}
& 0.2595 & 0.3101 & \underline{0.2848} & 0.0381 & 0.0825 & 0.8777 & \underline{0.3328} \\
Scenario 3
& 0.4167 & 0.5670 & \underline{0.4918} & 0.1042 & 0.3182 & 0.7807 & \underline{0.4010}
& 0.2658 & 0.2883 & \underline{0.2770} & 0.0313 & 0.0924 & 0.8871 & \underline{0.3369} \\
Scenario 4 
& 0.4479 & 0.5309 & \underline{0.4894} & 0.3229 & 0.4124 & 0.6930 & \underline{0.4761}
& 0.3034 & 0.2966 & \underline{0.3000} & 0.1682 & 0.1870 & 0.8419 & \underline{0.3990} \\
\textbf{Our Method} 
& \textbf{0.5104} & \textbf{0.5619} & \underline{\textbf{0.5362}} & \textbf{0.3229} & \textbf{0.4709} & \textbf{0.7368} & \underline{\textbf{0.5102}} 
& \textbf{0.3091} & \textbf{0.3185} & \underline{\textbf{0.3138}} & \textbf{0.2106} & \textbf{0.2556} & \textbf{0.8576} & \underline{\textbf{0.4413}} \\
\hline
\end{tabular}
\end{center}
\normalsize
\vspace{-3mm}
\end{table*}
\section{Experiments and Analysis}\label{sec-III}
\subsection{Dataset}
We design our continual procedure using three datasets to simulate a realistic VQA CL scenario, including function-incremental learning with a broader range of questions and answers, and scene-incremental learning with a greater variety of surgical types: we train our CL model with EndoVis17 at $t=1$, and then with EndoVis18 at $t=2$, and finally with DAISI-VQA at $t=3$. The first two datasets are well-known open-source datasets used in previous papers \cite{Ref038, Ref039, Ref040}, while the third one is a new dataset we developed with the assistance of GPT-3.5. Due to space limit, we do not present detailed introduction of the previous two datasets. We refer readers to our technical reports \cite{Ref000} for a comprehensive dataset description. We now introduce the new dataset as follows.




\textbf{DAISI-VQA} is a new VQA dataset we developed using the DAISI dataset \cite{Ref021}. The original DAISI dataset contains images and instructional texts of various surgical procedures on different organs. Each procedure consists of multiple images with corresponding instructional texts. We first clean the original DAISI dataset by deleting irrelevant images (say frames not containing any surgical contents) and unimportant descriptions (say those describing the hospital or the surgeon). We then generate QA pairs according to the text description for each image. Eventually, we obtained a new VQA dataset referred to as DAISI-VQA. Apart from expanding the answer sets for the questions related to action and organ, we add instrument-related questions in DAISI-VQA to introduce a new knowledge domain. There are 353 surgical images and 545 QA pairs in the new dataset. We assign around 80\% of the data for training and use the rest for testing. To generate reasonable questions and reliable answers for each image, we process the text description of each image with GPT-3.5 and apply the ICL technique. Due to space limit, we refer readers to our technical report \cite{Ref000} for implementation details of ICL.


\begin{figure*}[htbp]
  \centering
  \subfloat[]  {\label{plot7a} \includegraphics[width=0.283\textwidth]{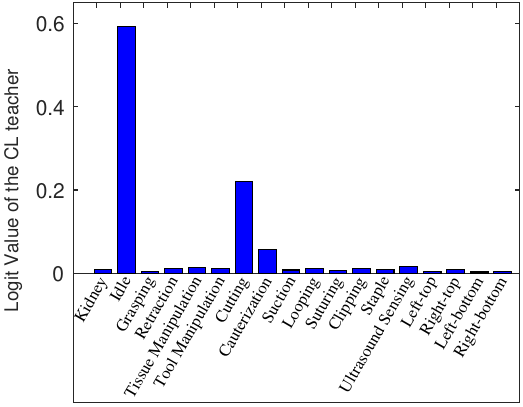}}
  \subfloat[]  {\label{plot7b} \includegraphics[width=0.283\textwidth]{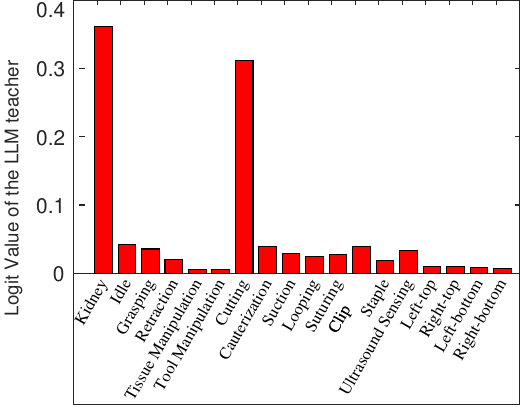}}
  \subfloat[]  {\label{plot7c} \includegraphics[width=0.283\textwidth]{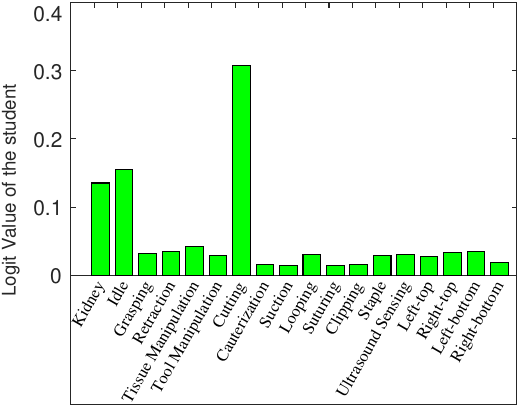}}
  \captionsetup{font={small}}
  \caption{Logits of CL teacher, LLM teacher, and the student for the case in Fig. \ref{Fig_6}.}
  \vspace{-1.5mm} 
  \label{plot7}
\end{figure*}

\subsection{Baselines and Implementation Details}
We evaluate our approach in comparison to the following algorithms under a CL setting: Fine Tune (FT) \cite{Ref025}, Experience Replay (ER) \cite{Ref042}, Learning without Forgetting (LwF) \cite{Ref001}, EWC++ \cite{Ref023}, and Online-EWC \cite{Ref022}. Due to space limit, we refer readers to our full-version technical report \cite{Ref000} for detailed discussions and comparisions between these baselines. 

For our method, we train the model on EndoVis17, EndoVis18, and DAISI-VQA at $t=1$, $t=2$, and $t=3$, respectively. For more implementation details, including experiment setups (e.g., learning rate and number of epochs) and training platform, we also refer readers to our technical report \cite{Ref000}.

\subsection{Experimental Results}
Before a detailed analysis, it is important to clarify that our comparison focuses on a model's \textit{average} performance (specifically, accuracy and F-score) across all datasets tested at different times, rather than its performance on an individual dataset. This is because a bad CL model suffering from catastrophic forgetting might perform well on one dataset but falter on previously encountered ones. Conversely, a robust CL model maintains consistently high performance across all datasets. Although a good CL model may not outshine a bad one in a singular dataset, it demonstrates superior performance on average. In light of this, comparing average performance across all tested datasets is a tradition in CL studies \cite{RefA2, RefA1, Ref041, Ref006}, and this paper follows the same setting.

We now take a look at the benchmarking results depicted in Table I. FT meets the severest catastrophic forgetting among all tested methods and gets the worst performance. Despite ER, the ideal upper bound, our technique consistently displays superior performance throughout the process. Compared with previous methods, our method improves the model accuracy of the second-best model by 9.56\% at $t=2$ and 7.17\% at $t=3$. We improve the F-score of the second-best model by 5.64\% and 10.58\% at $t=2$ and $t=3$, respectively.

The above results prove our method's remarkable ability in learning new knowledge without forgetting the old one. We attribute the good performance of the model to 1) the application of LLM in the training process for overcoming domain shift and data imbalance, and 2) the dynamic weight adjustment scheme that makes a good balance between the expert teacher model and the LLM. To demonstrate the efficacy of each component in our method, we carry out an ablation study with consideration of the following scenarios: \textbf{1)} in weight assignment, ignore data imbalance and consider domain shift only; \textbf{2)} in weight assignment, ignore domain shift and consider data imbalance only; \textbf{3)} remove the entire adaptive weight assignment mechanism, opting instead for a fixed weight in the CL process; and \textbf{4)} remove the LLM teacher and use the old CL teacher only.

Ablation study results are detailed in Table II. Our method beats its ablations in Scenario 1/2/3/4, which underscores that each component we propose is integral to the ultimate performance, hence proving themselves to be indispensable.

\section{Why LLM Works: Discussion and Case Study} \label{sec-IV}
A question one may ask is why using InstructBLIP as the additional teacher can significantly enhance the performance of the VQA task. Before answering the question, we want to point out that the multimodal LLM itself does not perform well when it tries to handle the surgical VQA task independently. In our experiments, we found that InstructBLIP tends to answer the task with "kidney". This is understandable, as InstructBLIP is trained on general domain datasets, which means the model is more familiar with daily used words (e.g., kidney) than other surgical terms. However, we emphasize that \textit{having poor performance when independently dealing with the surgical task does not mean few contributions under our multi-teacher CL framework}, as the information hidden in the LLM's logits is also vital. The rest of this section illustrates how the LLM helps the CL training process with a case study.

\begin{figure}[htbp]
  \centering
  \subfloat[]  {\label{Fig_5} \includegraphics[width=0.24\textwidth]{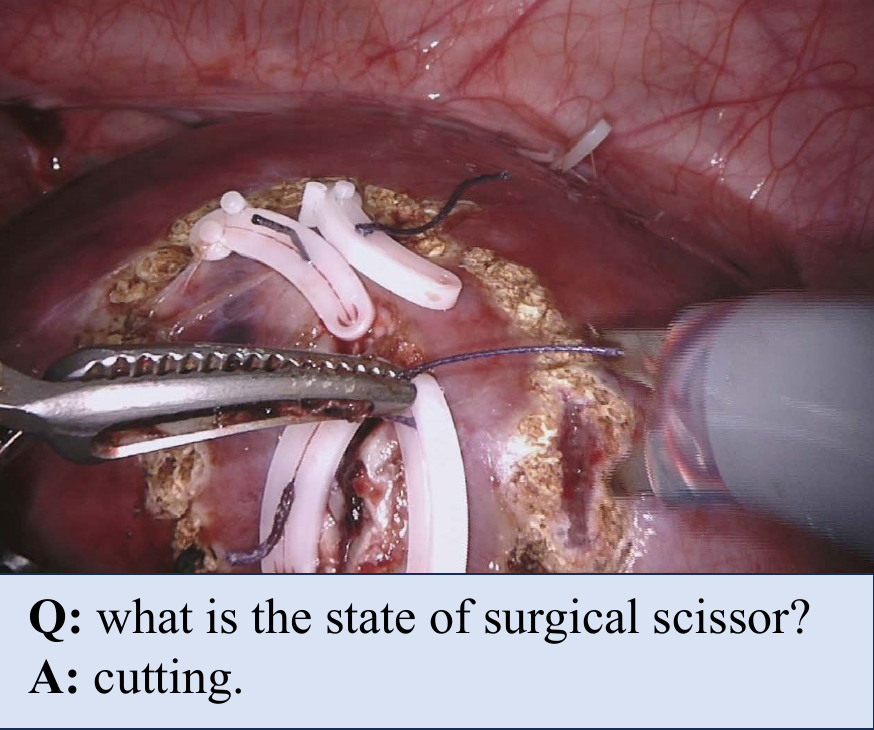}}
  \subfloat[]  {\label{Fig_6} \includegraphics[width=0.24\textwidth]{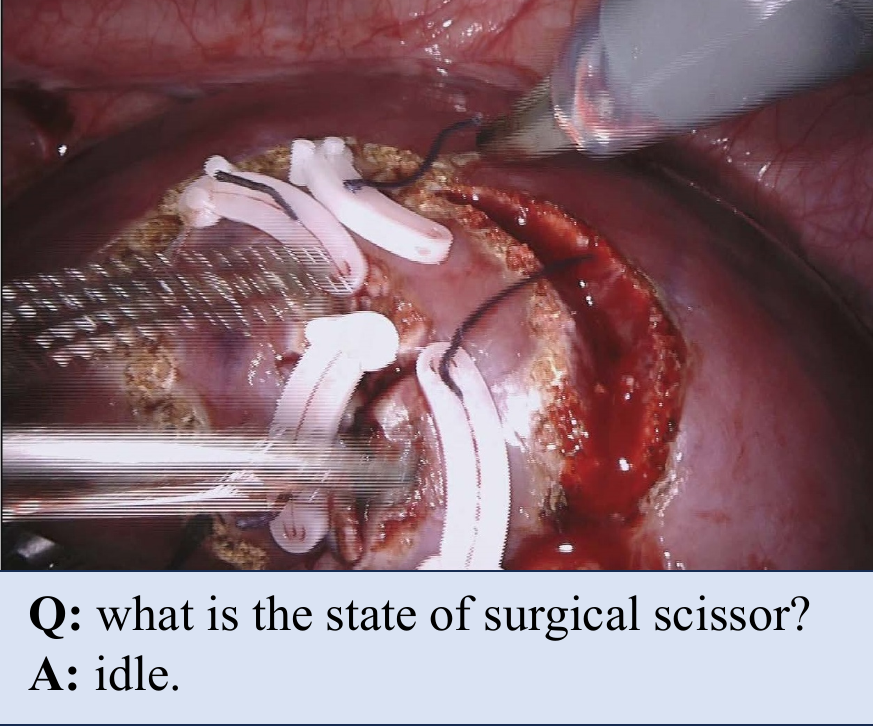}}
  \captionsetup{font={small}}
  \caption{Two EndoVis18 images showing the cutting operation on a kidney. If a conventional CL approach were applied, the VQA model at $t=2$ would misclassify the second operation in (b) as ``idle".}
  \vspace{-1mm}
  \label{plot5}
\end{figure}

At $t=2$, experimental results show that the CL teacher model (trained at $t=1$ on EndoVis17) often confuses the ``cutting" operation with the ``idle" operation, leading to poor training outcomes when a conventional single-teacher framework is used. The ``cutting" operation typically involves a surgical scissor. When the scissor is widely opened (Fig. \ref{Fig_5}), the CL teacher correctly classifies the operation as ``cutting". However, when the scissor is not significantly opened (Fig. \ref{Fig_6}), the CL teacher may misclassify the "cutting" operation as "idle". This confusion is understandable, as a slightly opened scissor during "cutting" may resemble an "idle" one. However, a senior surgeon could clearly identify the operation as "cutting" based on the partially cut organ shown in Fig. \ref{Fig_6}.

With our multi-teacher framework that uses LLM as a second teacher, the student model can effectively distinguish between ``cutting" and ``idle" operations. To understand how the LLM aids in the correct classification at $t=2$, we examine the logits of both the conventional teacher and the LLM teacher during the learning process. Fig. \ref{plot7a} and Fig. \ref{plot7b} depict the logits of the conventional CL teacher and that of the LLM teacher for the studied ``cutting" case. The old CL teacher assigns the highest logit value to ``idle" and the second highest to ``cutting", indicating that the confusion between the two operations does exist. The LLM teacher attributes the highest logits to ``kidney" and the second highest to ``cutting".
  
As the student's logits presented in Fig. \ref{plot7c}, although neither the conventional CL model nor the LLM teacher can independently generate the correct answer, their combined decision made by the weighted sum-up can correctly identify the operation as ``cutting". This observation highlights the importance of our proposed multi-teacher CL framework with adaptive weights, which effectively utilizes the hidden information of the LLM's logits to boost performance.

\section{Conclusion and Future Work}\label{sec-V}
This paper puts forth a novel methodology to enhance the performance of surgical VQA tasks for robotic-assisted surgical education under a CL setup. Through the proposed multi-teacher CL framework that utilizes a multimodal LLM to bridge the knowledge gap when facing domain shift and data imbalance, we effectively solve the catastrophic forgetting problem. Our innovative data processing scheme provides a method to extract logits from complex LLM embeddings, and our adaptive weight assignment approach ensures a balance between the domain-specific expertise of the old CL model and the strong generalization ability of the LLM. The application of these methods demonstrates compelling performance in tackling realistic surgical VQA tasks. Additionally, our new surgical VQA dataset offers a valuable resource for future research. Importantly, this paper puts forth a new research direction for leveraging LLMs in CL studies. The potential for improving continual learning VQA for clinical use remains significant. For future work, we will investigate how to decompose the representation into spatial space and temporal space, which has higher task-invariance, to further alleviate the model forgetting problem. Another potential direction is to integrate the multi-modal data to advance the performance outcomes. 

\bibliographystyle{IEEEtran}
\bibliography{ref}

\begin{thebibliography}{10}
\providecommand{\url}[1]{#1}
\csname url@samestyle\endcsname
\providecommand{\newblock}{\relax}
\providecommand{\bibinfo}[2]{#2}
\providecommand{\BIBentrySTDinterwordspacing}{\spaceskip=0pt\relax}
\providecommand{\BIBentryALTinterwordstretchfactor}{4}
\providecommand{\BIBentryALTinterwordspacing}{\spaceskip=\fontdimen2\font plus
\BIBentryALTinterwordstretchfactor\fontdimen3\font minus \fontdimen4\font\relax}
\providecommand{\BIBforeignlanguage}[2]{{%
\expandafter\ifx\csname l@#1\endcsname\relax
\typeout{** WARNING: IEEEtran.bst: No hyphenation pattern has been}%
\typeout{** loaded for the language `#1'. Using the pattern for}%
\typeout{** the default language instead.}%
\else
\language=\csname l@#1\endcsname
\fi
#2}}
\providecommand{\BIBdecl}{\relax}
\BIBdecl

\bibitem{Ref003}
H.~Wang, Y.~Jin, and L.~Zhu, ``Dynamic interactive relation capturing via scene graph learning for robotic surgical report generation,'' in \emph{2023 IEEE ICRA}, 2023, pp. 2702--2709.

\bibitem{Ref005}
Z.~Zhao, Y.~Jin, B.~Lu, C.-F. Ng, Q.~Dou, Y.-H. Liu, and P.-A. Heng, ``One to many: Adaptive instrument segmentation via meta learning and dynamic online adaptation in robotic surgical video,'' in \emph{2021 IEEE ICRA}, 2021, pp. 13\,553--13\,559.

\bibitem{Ref045}
Y.~Jin, K.~Cheng, Q.~Dou, and P.-A. Heng, ``Incorporating temporal prior from motion flow for instrument segmentation in minimally invasive surgery video,'' in \emph{Medical Image Computing and Computer Assisted Intervention (MICCAI) 2019}.\hskip 1em plus 0.5em minus 0.4em\relax Springer, 2019, pp. 440--448.

\bibitem{Ref006}
L.~Seenivasan, M.~Islam, A.~K. Krishna, and H.~Ren, ``{Surgical-VQA}: Visual question answering in surgical scenes using transformer,'' in \emph{2022 MICCAI}.\hskip 1em plus 0.5em minus 0.4em\relax Springer, 2022, pp. 33--43.

\bibitem{Ref007}
L.~Bai, M.~Islam, and H.~Ren, ``Co-attention gated vision-language embedding for visual question localized-answering in robotic surgery,'' \emph{arXiv preprint arXiv:2307.05182}, 2023.

\bibitem{Ref008}
L.~Seenivasan, M.~Islam, G.~Kannan, and H.~Ren, ``{SurgicalGPT}: End-to-end language-vision {GPT} for visual question answering in surgery,'' \emph{arXiv preprint arXiv:2304.09974}, 2023.

\bibitem{Ref035}
B.~D. Nguyen, T.-T. Do, B.~X. Nguyen, T.~Do, E.~Tjiputra, and Q.~D. Tran, ``Overcoming data limitation in medical visual question answering,'' in \emph{2019 MICCAI}.\hskip 1em plus 0.5em minus 0.4em\relax Springer, 2019, pp. 522--530.

\bibitem{Ref036}
Q.~Wu, P.~Wang, X.~Wang, X.~He, and W.~Zhu, ``Medical {VQA},'' in \emph{Visual Question Answering: From Theory to Application}.\hskip 1em plus 0.5em minus 0.4em\relax Springer, 2022, pp. 165--176.

\bibitem{Ref037}
Y.~Khare, V.~Bagal, M.~Mathew, A.~Devi, U.~D. Priyakumar, and C.~Jawahar, ``Mmbert: Multimodal {Bert} pretraining for improved medical vqa,'' in \emph{2021 ISBI}.\hskip 1em plus 0.5em minus 0.4em\relax IEEE, 2021, pp. 1033--1036.

\bibitem{Ref009}
R.~M. French, ``Catastrophic forgetting in connectionist networks,'' \emph{Trends in cognitive sciences}, vol.~3, no.~4, pp. 128--135, 1999.

\bibitem{RefA2}
K.~Shu, H.~Li, J.~Cheng, Q.~Guo, L.~Leng, J.~Liao, Y.~Hu, and J.~Liu, ``Replay-oriented gradient projection memory for continual learning in medical scenarios,'' in \emph{2022 IEEE BIBM}, 2022, pp. 1724--1729.

\bibitem{RefA1}
M.~M. Derakhshani, I.~Najdenkoska, T.~van Sonsbeek, X.~Zhen, D.~Mahapatra, M.~Worring, and C.~G. Snoek, ``Lifelonger: A benchmark for continual disease classification,'' in \emph{2022 MICCAI}.\hskip 1em plus 0.5em minus 0.4em\relax Springer, 2022, pp. 314--324.

\bibitem{RefA4}
M.~Lenga, H.~Schulz, and A.~Saalbach, ``Continual learning for domain adaptation in chest x-ray classification,'' in \emph{Medical Imaging with Deep Learning}.\hskip 1em plus 0.5em minus 0.4em\relax PMLR, 2020, pp. 413--423.

\bibitem{Ref001}
Z.~Li and D.~Hoiem, ``Learning without forgetting,'' \emph{IEEE Trans. Pattern Anal. Mach. Intell.}, vol.~40, no.~12, pp. 2935--2947, 2017.

\bibitem{Ref002}
J.~Kirkpatrick, R.~Pascanu, N.~Rabinowitz, J.~Veness, G.~Desjardins, A.~A. Rusu, K.~Milan, J.~Quan, T.~Ramalho, A.~Grabska-Barwinska \emph{et~al.}, ``Overcoming catastrophic forgetting in neural networks,'' \emph{Proc. Nat. Acad. Sci.}, vol. 114, no.~13, pp. 3521--3526, 2017.

\bibitem{RefA3}
L.~Bai, M.~Islam, and H.~Ren, ``Revisiting distillation for continual learning on visual question localized-answering in robotic surgery,'' \emph{arXiv preprint arXiv:2307.12045}, 2023.

\bibitem{Ref027}
M.~Phuong and C.~Lampert, ``Towards understanding knowledge distillation,'' in \emph{2019 ICML}.\hskip 1em plus 0.5em minus 0.4em\relax PMLR, 2019, pp. 5142--5151.

\bibitem{Ref028}
C.~Simon, M.~Faraki, Y.-H. Tsai, X.~Yu, S.~Schulter, Y.~Suh, M.~Harandi, and M.~Chandraker, ``On generalizing beyond domains in cross-domain continual learning,'' in \emph{2022 CVPR}, 2022, pp. 9265--9274.

\bibitem{Ref029}
T.~Diethe, T.~Borchert, E.~Thereska, B.~Balle, and N.~Lawrence, ``Continual learning in practice,'' \emph{arXiv preprint arXiv:1903.05202}, 2019.

\bibitem{Ref030}
C.~D. Kim, J.~Jeong, and G.~Kim, ``Imbalanced continual learning with partitioning reservoir sampling,'' in \emph{2020 ECCV}.\hskip 1em plus 0.5em minus 0.4em\relax Springer, 2020, pp. 411--428.

\bibitem{Ref031}
A.~Chrysakis and M.-F. Moens, ``Online continual learning from imbalanced data,'' in \emph{2020 ICML}.\hskip 1em plus 0.5em minus 0.4em\relax PMLR, 2020, pp. 1952--1961.

\bibitem{Ref010}
A.~Belyaeva, J.~Cosentino, F.~Hormozdiari, C.~Y. McLean, and N.~A. Furlotte, ``Multimodal {LLMs} for health grounded in individual-specific data,'' \emph{arXiv preprint arXiv:2307.09018}, 2023.

\bibitem{Ref011}
Y.~Du, S.~C. Liew, K.~Chen, and Y.~Shao, ``The power of large language models for wireless communication system development: A case study on {FPGA} platforms,'' \emph{arXiv preprint arXiv:2307.07319}, 2023.

\bibitem{Ref012}
Z.~Guo, R.~Zhang, X.~Zhu, Y.~Tang, X.~Ma, J.~Han, K.~Chen, P.~Gao, X.~Li, H.~Li \emph{et~al.}, ``Point-bind \& point-{LLM}: Aligning point cloud with multi-modality for {3D} understanding, generation, and instruction following,'' \emph{arXiv preprint arXiv:2309.00615}, 2023.

\bibitem{Ref043}
H.~Cui, Y.~Du, Q.~Yang, Y.~Shao, and S.~C. Liew, ``{LLMind}: Orchestrating {AI} and {IoT} with {LLMs} for complex task execution,'' \emph{arXiv preprint arXiv:2312.09007}, 2023.

\bibitem{Ref013}
A.~J. Thirunavukarasu, D.~S.~J. Ting, K.~Elangovan, L.~Gutierrez, T.~F. Tan, and D.~S.~W. Ting, ``Large language models in medicine,'' \emph{Nature medicine}, pp. 1--11, 2023.

\bibitem{Ref014}
H.~Nori, N.~King, S.~M. McKinney, D.~Carignan, and E.~Horvitz, ``Capabilities of {GPT-4} on medical challenge problems,'' \emph{arXiv preprint arXiv:2303.13375}, 2023.

\bibitem{Ref015}
S.~Min, X.~Lyu, A.~Holtzman, M.~Artetxe, M.~Lewis, H.~Hajishirzi, and L.~Zettlemoyer, ``Rethinking the role of demonstrations: What makes in-context learning work?'' \emph{arXiv preprint arXiv:2202.12837}, 2022.

\bibitem{Ref032}
S.~Min, M.~Lewis, L.~Zettlemoyer, and H.~Hajishirzi, ``Metaicl: Learning to learn in context,'' \emph{arXiv preprint arXiv:2110.15943}, 2021.

\bibitem{Ref044a}
K.~Chen, J.~Li, K.~Wang, Y.~Du, J.~Yu, J.~Lu, G.~Chen, L.~Li, J.~Qiu, Q.~Fang \emph{et~al.}, ``Towards an automatic {AI} agent for reaction condition recommendation in chemical synthesis,'' \emph{arXiv preprint arXiv:2311.10776}, 2023.

\bibitem{Ref044b}
K.~Chen, H.~Cao, J.~Li, Y.~Du, M.~Guo, X.~Zeng, L.~Li, J.~Qiu, P.~A. Heng, and G.~Chen, ``An autonomous large language model agent for chemical literature data mining,'' \emph{arXiv preprint arXiv:2402.12993}, 2024.

\bibitem{Ref033}
O.~Rubin, J.~Herzig, and J.~Berant, ``Learning to retrieve prompts for in-context learning,'' \emph{arXiv preprint arXiv:2112.08633}, 2021.

\bibitem{Ref026}
J.~Gou, B.~Yu, S.~J. Maybank, and D.~Tao, ``Knowledge distillation: A survey,'' \emph{International Journal of Computer Vision}, vol. 129, pp. 1789--1819, 2021.

\bibitem{Ref034}
X.~Dai, Z.~Jiang, Z.~Wu, Y.~Bao, Z.~Wang, S.~Liu, and E.~Zhou, ``General instance distillation for object detection,'' in \emph{2021 CVPR}, 2021, pp. 7842--7851.

\bibitem{Ref016}
D.~Wenliang, L.~Junnan, L.~Dongxu, T.~Anthony Meng~Huat, Z.~Junqi, W.~Weisheng, L.~Boyang, F.~Pascale, and H.~Steven, ``{InstructBLIP}: Towards general-purpose vision-language models with instruction tuning,'' \emph{arXiv preprint arXiv:2305.06500}, 2023.

\bibitem{Ref017}
H.~W. Chung, L.~Hou, S.~Longpre, B.~Zoph, Y.~Tay, W.~Fedus, E.~Li, X.~Wang, M.~Dehghani, S.~Brahma \emph{et~al.}, ``Scaling instruction-finetuned language models,'' \emph{arXiv preprint arXiv:2210.11416}, 2022.

\bibitem{Ref018}
F.~Thabtah, S.~Hammoud, F.~Kamalov, and A.~Gonsalves, ``Data imbalance in classification: Experimental evaluation,'' \emph{Information Sciences}, vol. 513, pp. 429--441, 2020.

\bibitem{Ref038}
L.~Maier-Hein, M.~Eisenmann, D.~Sarikaya, K.~M{\"a}rz, T.~Collins, A.~Malpani, J.~Fallert, H.~Feussner, S.~Giannarou, P.~Mascagni \emph{et~al.}, ``Surgical data science--from concepts toward clinical translation,'' \emph{Medical image analysis}, vol.~76, p. 102306, 2022.

\bibitem{Ref039}
Z.-L. Ni, G.-B. Bian, Z.-G. Hou, X.-H. Zhou, X.-L. Xie, and Z.~Li, ``Attention-guided lightweight network for real-time segmentation of robotic surgical instruments,'' in \emph{2020 ICRA}.\hskip 1em plus 0.5em minus 0.4em\relax IEEE, 2020, pp. 9939--9945.

\bibitem{Ref040}
J.~Liu, X.~Guo, and Y.~Yuan, ``Graph-based surgical instrument adaptive segmentation via domain-common knowledge,'' \emph{IEEE Trans. Med. Imaging}, vol.~41, no.~3, pp. 715--726, 2021.

\bibitem{Ref000}
K.~Chen, Y.~Du, T.~You, M.~Islam, Z.~Guo, Y.~Jin, G.~Chen, and P.~A. Heng, ``{LLM}-assisted multi-teacher continual learning for visual question answering in robotic surgery,'' \emph{arXiv preprint arXiv:2402.16664}, 2024.

\bibitem{Ref021}
E.~Rojas-Mu{\~n}oz, K.~Couperus, and J.~Wachs, ``Daisi: database for {AI} surgical instruction,'' \emph{arXiv preprint arXiv:2004.02809}, 2020.

\bibitem{Ref025}
S.~W. Lei, D.~Gao, J.~Z. Wu, Y.~Wang, W.~Liu, M.~Zhang, and M.~Z. Shou, ``Symbolic replay: Scene graph as prompt for continual learning on vqa task,'' in \emph{2023 AAAI}, vol.~37, no.~1, 2023, pp. 1250--1259.

\bibitem{Ref042}
S.~Zhang and R.~S. Sutton, ``A deeper look at experience replay,'' \emph{arXiv preprint arXiv:1712.01275}, 2017.

\bibitem{Ref023}
A.~Chaudhry, P.~K. Dokania, T.~Ajanthan, and P.~H. Torr, ``Riemannian walk for incremental learning: Understanding forgetting and intransigence,'' in \emph{2018 ECCV}, 2018, pp. 532--547.

\bibitem{Ref022}
F.~Husz{\'a}r, ``Note on the quadratic penalties in elastic weight consolidation,'' \emph{Proc. Nat. Acad. Sci.}, vol. 115, no.~11, pp. E2496--E2497, 2018.

\bibitem{Ref041}
L.~Bai, M.~Islam, L.~Seenivasan, and H.~Ren, ``Surgical-{VQLA}: Transformer with gated vision-language embedding for visual question localized-answering in robotic surgery,'' \emph{arXiv preprint arXiv:2305.11692}, 2023.

\end{thebibliography}
\end{document}